\documentclass[conference]{IEEEtran}
\usepackage{cite}
\usepackage{amsmath,amssymb,amsfonts}
\usepackage{listings}
\usepackage{algorithm}
\usepackage{algpseudocode}
\usepackage{graphicx}
\usepackage{textcomp}
\usepackage{multirow}
\usepackage{xcolor}
\usepackage{soul}
\usepackage{url}
\def\BibTeX{{\rm B\kern-.05em{\sc i\kern-.025em b}\kern-.08em
    T\kern-.1667em\lower.7ex\hbox{E}\kern-.125emX}}

\makeatletter
\def\endthebibliography{%
  \def\@noitemerr{\@latex@warning{Empty `thebibliography' environment}}%
  \endlist
}
\makeatother

\DeclareMathOperator*{\sumi}{\text{\Huge $\Xi$}}

\begin{document}

\title{Exploring FPGA designs for MX and beyond}

\author{\IEEEauthorblockN{Ebby Samson}
\IEEEauthorblockA{\textit{Imperial College London}\\
London, UK \\
es1219@ic.ac.uk}
\and
\IEEEauthorblockN{Naveen Mellempudi}
\IEEEauthorblockA{\textit{AMD}\\
Austin, USA \\
naveen.mellempudi@amd.com}
\and
\IEEEauthorblockN{Wayne Luk}
\IEEEauthorblockA{\textit{Imperial College London}\\
London, UK \\
w.luk@imperial.ac.uk}
\and
\IEEEauthorblockN{George A. Constantinides}
\IEEEauthorblockA{\textit{Imperial College London}\\
London, UK \\
g.constantinides@imperial.ac.uk}
}

\maketitle

\begin{abstract}
A number of companies recently worked together to release the new Open Compute Project MX standard for low-precision computation, aimed at efficient neural network implementation. In this paper, we describe and evaluate the first open-source FPGA implementation of the arithmetic defined in the standard. Our designs fully support all the standard's concrete formats for conversion into and out of MX formats and for the standard-defined arithmetic operations, as well as arbitrary fixed-point and floating-point formats. Certain elements of the standard are left as implementation-defined, and we present the first concrete FPGA-inspired choices for these elements, which we outline in the paper. Our library of optimized hardware components is available open source, and can be used to build larger systems. For this purpose, we also describe and release an open-source Pytorch library for quantization into the new standard, integrated with the Brevitas library so that the community can develop novel neural network designs quantized with MX formats in mind. We demonstrate the usability and efficacy of our libraries via the implementation of example neural networks such as ResNet-18 on the ImageNet ILSVRC12 dataset. Our testing shows that MX is very effective for formats such as INT5 or FP6 which are not natively supported on GPUs. This gives FPGAs an advantage as they have the flexibility to implement a custom datapath and take advantage of the smaller area footprints offered by these formats.

\end{abstract}


\section{Introduction}

Developments in digital hardware over the past few decades have greatly increased the compute performance available for machine learning training and inference. This has allowed larger neural networks with more parameters to be trained and deployed, yielding ever more powerful models with better accuracy. However, the demand for more computational power is still increasing with demand for models with more parameters~\cite{Kaplan2020ScalingLF}. Aside from this, there is demand for the ability to deploy powerful models with large numbers of parameters in low-power environments such as edge devices. For these reasons, lots of research in digital hardware design had the goal of creating hardware which allows deploying these models with smaller footprints while preserving inference accuracy.

Traditionally, machine learning inference and training are performed using the IEEE FP32 format, with all values such as parameters, activations, gradients and weight updates represented in FP32. However, there has been a growing demand for compact data representations with high-throughput compute. The MX standard~\cite{ocp_mx} aims to help create more compact models while preserving the accuracy of their full-precision counterparts. The standard introduces quantization with a new scale sharing regime, as compared to traditional per-tensor or per-channel quantization. Our contributions are the following:

\begin{itemize}
    \item We present the first open-source FPGA-oriented implementation of the new MX standard.
    \item We explore parameter choices beyond the concrete implementations defined in the standard and evaluate their impact on inference accuracy and FPGA area, allowing us to uncover some promising design points.
    \item We provide open-source software infrastructure to facilitate exploration with MX and similar schemes in Pytorch.
\end{itemize}

\section{Background}

\subsection{Quantization}

\begin{table*}[t!]
\caption{Restrictions on the scales of a weight tensor ($F$) and activation tensor ($A$) imposed by scale sharing regimes on a 2D convolution example. The block size for MX scaling is denoted by $k$. $S$ and $T$ are scales for $F$ and $A$ respectively.}
\label{tab:scale_share}
\centering
\resizebox{0.7\textwidth}{!}{\begin{tabular}{ ||c||c|c|| } 
\hline
Scale Sharing  & \multicolumn{2}{c||}{Restrictions} \\ \hline
& Weights & Activations \\
\hline \hline
 \  & F, S $ \in\mathbb{R}^{K \times C \times H' \times W'} $ & A, T $ \in\mathbb{R}^{N \times H \times W \times C} $  \\
\hline

Per-Tensor  & $ \forall l, c, h', w' \ \ S_{l, c, h', w'} = s $ & $ \forall n, h, w, c \ \ T_{n, h, w, c} = t $ \\
\hline
Per-Channel  & $ \forall l, h', w' \ \ S_{l, c, h', w'} = s_{c} $ & $ \forall n, h, w \ \ T_{n, h, w, c} = t_{c} $ \\
\hline \hline
 \  & S $ \in\mathbb{R}^{K \times \lceil \frac{C}{k} \rceil \times H' \times W' \times k} $ & T $ \in\mathbb{R}^{N \times H \times W \times \lceil \frac{C}{k} \rceil \times k} $  \\
\hline
MX  & $ \forall p \ \ S_{l, c, h', w', p} = s_{l, c, h', w'} $ & $ \forall p \ \ T_{n, h, w, c, p} = t_{n, h, w, c} $ \\
\hline
\end{tabular}}
\end{table*}

One common approach to reducing area and bandwidth of neural networks is by exploring number representations where fewer bits are used per parameter than in IEEE 754~\cite{8766229} floating-point formats. A downside of these narrow formats is that they typically have lower precision or dynamic range in comparison. The reduction in dynamic range is extreme in floating-point formats with narrow exponents~\cite{8280150} and fixed-point formats~\cite{Courbariaux2014TrainingDN}. These formats are usually coupled with a shared scale factor and zero point~\cite{Nagel2021AWP} which transform values to make better use of the available range. This transformation is shown in Equation~\ref{eqn:quant} where tensors $S$ and $Z$ represent the scale and zero point respectively and $\odot$ represents element-wise multiplication. Equation~\ref{eqn:quant_mx} shows restrictions imposed by the MX standard, where $\mathbb{Z}_{b}$ denotes the set of values representable by a $b$-bit integer and $\mathbb{M}_{e,m}$ denotes the set of values representable by a floating-point format with an $e$-bit exponent and $m$-bit mantissa. Quantized tensor $X_{q}$ can be integer or floating-point. The scale and zero point scale and shift values such that most of them lie in the range representable by the target format, and that most of them have distinct values in the target format. They are usually shared with tensor-wise or channel-wise granularity due to the computational cost savings from factoring them out of dot products. Table~\ref{tab:scale_share} shows the restrictions imposed by scale sharing regimes on a 2D convolution (Equation~\ref{eqn:dp_conv}). Per-tensor scales restrict all elements of the scale to the same value, while per-channel scales allow scales to vary along the $C$ dimension. In the case of per-vector~\cite{Dai2021VSQuantPS} and MX~\cite{rouhani2023microscaling} scaling, the scale tensor is reshaped such that the $C$ dimension is replaced with two dimensions $\lceil \frac{C}{k} \rceil$ and $k$. In this configuration, only values along the $k$ dimension share a common scale. The need for reshaping will be explained in Section~\ref{sec:exp_infra_mf_quant}.

\begin{equation}
\label{eqn:quant}
\begin{aligned}
&X = S \odot (X_{q} - Z) \quad \text{where} \quad X\in\mathbb{R}^{d_1 \times d_2 \times \ldots \times d_n}
\end{aligned}
\end{equation}

\begin{equation}
\label{eqn:quant_mx}
\begin{gathered}
S\in\mathbb{M}_{8,0}^{d_1 \times d_2 \times \ldots \times d_n} \quad Z = 0 \\
X_{q} \in \mathbb{Z}_{b}^{d_1 \times d_2 \times \ldots \times d_n} \quad \text{ or } \quad X_{q} \in \mathbb{M}_{e,m}^{d_1 \times d_2 \times \ldots \times d_n} 
\end{gathered}
\end{equation}

\begin{multline}
\label{eqn:dp_conv}
\begin{aligned}
&A'_{n,h,w,l} = \sum_{c=1}^{C} \sum_{h'=1}^{H'} \sum_{w'=1}^{W'} (A(n, h+h', w+w', c) \\
&\quad \quad \quad \quad \quad \quad \quad \quad \quad \quad \quad \times F(l, c, h', w'))
\end{aligned}
\end{multline}

\subsection{Shared Scale Boundaries}

Traditionally, a scale value is shared between all elements in a tensor as this allows the scale to be factored out of dot products. This however limits all values in the tensor to be within a small range dictated by the dynamic range of the element format. Per-channel scales ease this by allowing each input channel in activation and weight tensors to have a different scale. Per-vector scales~\cite{Dai2021VSQuantPS} use even finer granularity. As granularity gets finer, quantization noise generally decreases. Another method proposes that tensors are divided into square blocks~\cite{BMF} as this keeps blocks contiguous during the backward pass. The MX standard~\cite{rouhani2023microscaling} is a special case of per-vector scaling with restrictions such as a power-of-two scale. The dimension used for sharing exponents in per-vector and MX scaling is called the {\em principal dimension}. A block refers to a group of elements of a tensor which share a scale value.

There are two main considerations when drawing block boundaries. First is the extra bandwidth required to load the scales along with parameters. Second is the effect on accumulation during dot products between the activation and weight tensors. Per-tensor scaling minimizes these effects by using fewer scales, thereby allowing accumulation to be fully performed with scales factored out. In other cases, accumulations appear across shared scale boundaries.

\subsection{The MX Standard}

The MX Standard~\cite{rouhani2023microscaling} is aimed at efficient neural network implementation by reducing the loss in accuracy when using fewer than 8 bits to represent parameters~\cite{Zhang_2023}. The standard uses a fine grained scale restricted to a ``E8M0" format (8-bit exponent, 0-bit mantissa), which restricts the scale to powers of two. The MX standard recommends that the scale is set to the largest power-of-two in the block, divided by the largest power-of-two representable in the element format. This presents new challenges in hardware implementation as converting into this format in hardware requires that area is dedicated to computing statistics over the input tensor. Normalising output activations during inference also requires computing statistics, this process recomputes the scale after the values in a block have been modified.

The standard introduces a set of concrete MX compliant formats shown in Table~\ref{tab:mx_concrete}. Concrete formats use a block size ($k$) of 32. The standard also defines arithmetic operations for computing dot products between vectors with MX scaling (Equation~\ref{eqn:std_ops}). This equation shows how a dot product is calculated between tensors $X$ and $Y$, which are associated with MX scales $S$ and $T$. The standard leaves some aspects of these operations as implementation-defined, such as the internal precision used for accumulation. In our implementation (Section~\ref{sec:impl}), we make use of this freedom to optimise for FPGA and test alternative choices where the standard allows, including alternative block sizes and element types.

\begin{table}[h!]
\centering
\caption{MX concrete formats and formats supported by our implementation. $e \in [2,6]$, $m \in [1,5]$, $b \in [2,8]$, $\log_2(k) \in [2,9]$. All scales are E8M0. Special behaviour follows OCP FP8~\cite{ocp_fp8}.}
\begin{tabular}{ |c|c|c|c| } 
\hline
Format Name & Element Type & Block Size & Specials/Behaviour \\
\hline
MXFP8  & E5M2 & 32 & NaN, Inf / OFL, SAT \\
\hline
MXFP8  & E4M3 & 32 & NaN / OFL, SAT  \\
\hline
MXFP6  & E3M2 & 32 & -  \\
\hline
MXFP6  & E2M3 & 32 & -  \\
\hline
MXFP4  & E2M1 & 32 & -  \\
\hline
MXINT8  & INT8 & 32 & -  \\
\hline
Our MXFP  & E$e$M$m$ & $k$ & NaN, Inf / OFL, SAT  \\
\hline
Our MXINT  & INT$b$ & $k$ & -  \\
\hline
\end{tabular}
\label{tab:mx_concrete}
\end{table}

\subsection{Brevitas}

Brevitas is a Pytorch library that facilitates quantization of neural networks. The library allows quantizers to be inserted into the compute graph of a model. As an example, they can quantize weights and activations before a convolution. The purpose of these quantizers is to model the effect of quantization in the model by injecting quantization noise. Quantizers use rounding and clipping to mimic the precision and range of a target format. This is done by passing inputs through a quantization function, followed a dequantization function which converts quantized tensors to FP32. This allows quantization-aware training (QAT) to be performed on FP32 hardware, which improves accuracy of the model on low-precision hardware. This also allows testing the accuracy impact of quantization while running on FP32 hardware. In this paper, we extend Brevitas with quantization to MX.

\begin{equation}
\label{eqn:std_ops}
\begin{aligned}
&\text{DotGeneral}(X, Y, S, T) = \sum_{c=1}^{C} \text{Dot}(X_c, Y_c, S_{c}, T_{c}) \\
&\text{Dot}(A, B, s, t) = (s t) \sum_{p=1}^{k} A_p B_p \\
& X,Y \in \mathbb{R}^{C \times k} \quad A, B \in \mathbb{R}^{k} \quad S, T \in \mathbb{R}^{C}
\end{aligned}
\end{equation}

\section{IP Cores} \label{sec:impl}

Our open-source IP cores~\cite{ip_cores} support all of the concrete formats introduced by the MX standard as well as the standard-defined arithmetic operations. The cores comply with the MX standard and present new choices for details which are left as implementation-defined by the standard, such as handling of special values and the internal precision of accumulation. For these elements, we made choices with the goal of efficient FPGA implementation. We also take advantage of the customisability of FPGAs to support arbitrary precision integer and floating-point formats. The components in our library can be used to implement the concrete MX formats by setting parameters. In addition, our blocks support other configurations of element types and block sizes within the constraints shown in Table~\ref{tab:mx_concrete}.

\subsection{Special Values}

The MX standard describes two ways of representing special values. In one case, the shared scale can be set to the NaN encoding (\verb|0xFF|) to set all elements associated with it to NaN. In the second case where the element format supports special encodings, individual elements can be set to specials. In order to comply with the MX standard, the FP8 concrete formats need to support the OCP FP8 standard~\cite{ocp_fp8} which describes special encodings and the behaviour of these specials in {\em overflow} and {\em saturating} modes. There are four different types of special behaviour depending on the FP8 format (E5M2 or E4M3) and the mode ({\em overflow} or {\em saturating}), as described in the OCP FP8 specicifation. Excluding FP8, none of the other concrete formats support special encodings in the element type, and the logic for propagating specials is left as implementation-defined by the standard. In our implementation, the four types of special behaviour from the OCP FP8 specification can be used with any MXFP format, or all special encodings can be disabled at the element level. In these cases where the element format does not support special encodings, if the input scale is a NaN, the output scale is set to the NaN encoding. All conversions propagate NaNs by setting output elements to the NaN encoding, either by setting an element to NaN if possible or by setting the shared scale to the NaN encoding if the element format does not support NaNs. Infs are propagated according to the OCP FP8 specification if the element format has an encoding for Inf, otherwise it is treated as a NaN. Specials are propagated similarly in dot product circuits.

\subsection{Dot Product Circuits}

\begin{figure}[htbp]
\centerline{\includegraphics[scale=0.6]{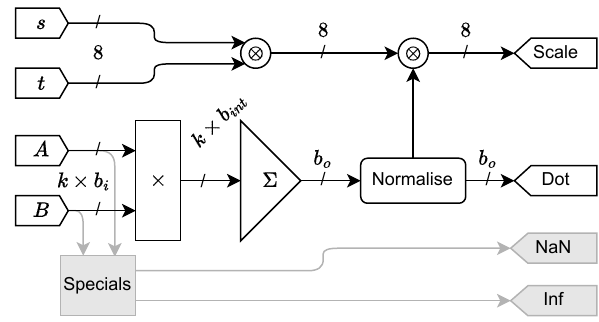}}
\caption{Our implementation of the {\em Dot} standard-defined operation, grey symbols are used for formats with special encodings. Table~\ref{tab:dp_widths} shows widths of signals. Multiplier inputs can be FP or INT but outputs are always INT.}
\label{fig:dp_circuits}
\end{figure}

The {\em Dot} standard-defined operation computes the dot product between a pair of blocks (Equation~\ref{eqn:std_ops}). The standard leaves the internal precision of this process as implementation-defined. As this operation factors out shared scales and is expected to be used with formats with small dynamic range, error-free integer accumulation is a viable option in hardware using Kulisch accumulation~\cite{Kulisch+2012}. Our implementation uses this error-free accumulation, and arranges adders in a binary tree structure to perform pairwise summation~\cite{doi:10.1137/1.9780898718027}, for low latency. Our implementation is shown in Figure~\ref{fig:dp_circuits}, and Table~\ref{tab:dp_widths} shows the internal precision used for floating-point and integer element types. The triangular block represents pairwise summation in hardware. If the element format has special encodings, the dot product circuit also checks input elements for specials, and sets the NaN or Inf flags.

\vspace{-0.1cm}

\begin{table}[ht!]
\centering
\caption{Dot product circuit configurations. $B$ denotes width of an integer, $E, M$ denote widths of the exponent and mantissa fields of a floating-point number.}
\begin{tabular}{ |c|c|c| } 
\hline
Width  & MXFP & MXINT  \\
\hline
$b_i$  & $1+E+M$ & $B$  \\
\hline
$b_{int}$  & $2(1+2^E+(M-1))$ & $2B$  \\
\hline
$b_o$  & $2(1+2^E+(M-1)) + \log_2(k)$ & $2B+\log_2(k)$  \\
\hline
\end{tabular}
\label{tab:dp_widths}
\end{table}

\begin{figure}[htbp]
\centerline{\includegraphics[scale=0.6]{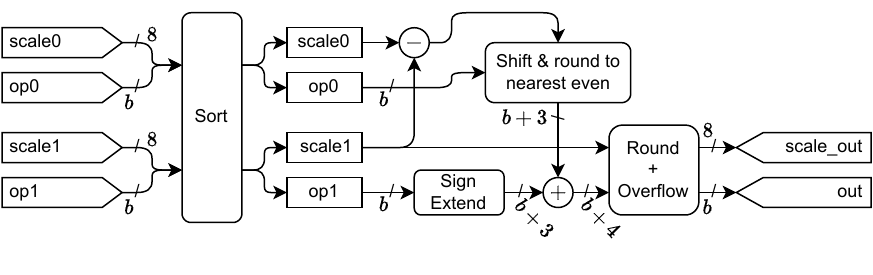}}
\caption{An adder that normalises operands, similar to a floating-point adder.}
\label{fig:add_nrm}
\end{figure}

While the {\em Dot} standard-defined operation performs dot products within the boundaries of a shared scale, the {\em Dot General} standard-defined operation performs dot products across boundaries. Our implementation of this operation replicates the {\em Dot} operation for each block within the input vector, then accumulates the outputs of the {\em Dot} operations using adders with normalisation. The standard does not define the internal precision to be used for this accumulation with normalisation, so our implementation uses a component similar to a floating-point adder as shown in Figure~\ref{fig:add_nrm}. This adder uses "round to nearest even" where precision is lost, the label $b+3$ in the figure represents the bit width of the input concatenated with the guard, round and sticky bits used for rounding.

\subsection{Conversion and Normalisation}

Our open-source library provides conversions from IEEE 754 FP32/BFloat16 and back, to facilitate integration of MX arithmetic into existing designs. The BFloat16 to MX converter can also normalise outputs of dot product circuits. The converters take a block of floating-point values along with their pre-computed scale. This is used to convert the input to a block of MX values, using "round to nearest even".

The MX standard recommends computing the scale during inference, while most previous work computes the scale during training, keeping it constant during inference~\cite{DBLP:journals/corr/abs-1903-08066}~\cite{Choi2018PACTPC}. The method recommended by the standard requires extra area. In particular, area is required to compute statistics during normalisation after dot product operations. Normalisation will also require variable shifts rather than constant shifts because scales are no longer constant at inference. Our IP cores can be used with constant pre-computed scales if desired, but our library also provides a component to compute the scale, implemented using a tree of comparators.

The block size used by all concrete formats is ($k=32$). Our IP cores implement the block size as a parameter which can be modified and set to any power of two according to the constraint in Table~\ref{tab:mx_concrete}. If the block size of the converter is modified, the depth of the comparator tree is set to to $\lceil \log_2(k) \rceil$ and new pipeline stages are added to keep the critical path two comparators long. This preserves timing characteristics.

\section{Exploration Infrastructure} \label{sec:explr_infra}

To facilitate developing models for the MX standard, we developed an open-source infrastructure~\cite{brevitas_mx} to allow Quantization Aware Training (QAT) and evaluation of MX quantized models on GPU. To allow mixing MX with existing quantization schemes, this infrastructure has been integrated into Brevitas~\cite{brevitas}. Integration with Brevitas gives the ability to change any aspect of quantization schemes and conduct design space exploration on Pytorch models. It also allows QAT with a quantized forward pass and FP32 backward pass.

The MX standard was released along with a Pytorch library that allows exploration of MX formats~\cite{mx_emu}. Our infrastructure expands the exploration space to provide many more choices such as scale types and other scale computation methods, including choices not supported by the MX standard such as floating-point scales. In addition, our infrastructure allows mixing quantization schemes such as MX and per-tensor.

\subsection{Minifloat Quantization} \label{sec:exp_infra_mf_quant}

Most MX concrete formats use floating-point. At the time of writing, the latest version of Brevitas~\cite{brevitas} does not support quantization schemes with low-precision floating-point elements. As a result, we have developed quantizers to floating-point element types and integrated these into Brevitas to compliment the existing fixed-point, binary and ternary types.

Brevitas supports several ways to collect statistics and compute scales. The method recommended by the MX standard sets the scale to the largest power-of-two present in the input divided by the largest power-of-two representable in the target format, and is already in Brevitas. Other Brevitas methods use the mean or a histogram when calculating the scale. These were designed for per-tensor and per-channel quantization where statistics are calculated along a set of dimensions. As MX could have multiple blocks in a single dimension, an input tensor needs to be reshaped before statistics can be computed (Equation~\ref{eqn:stats_reshape}). The {\em principal dimension} (input channels) is split into $\lceil \frac{C}{k} \rceil$ blocks of $k$. This reshaping is paired with zero padding such that $C \mod k = 0$ and allows statistics to be computed along the innermost dimension, ensuring compatibility with existing Brevitas scale implementations.

\begin{equation}
\label{eqn:stats_reshape}
    S \in\mathbb{R}^{K \times C \times H' \times W'} \ \  \text{to}  \ \   S \in\mathbb{R}^{K \times \lceil \frac{C}{k} \rceil \times H' \times W' \times k}
\end{equation}

The MX standard recommends that the scale is computed during inference. All current Brevitas methods compute a scale during training and keep it constant during inference, as this is what most previous work has used~\cite{DBLP:journals/corr/abs-1903-08066}~\cite{Choi2018PACTPC}. Using the standard's recommendation will add extra area to compute the scale during normalisation. Our exploration infrastructure allows the choice between the standard's recommended method and a constant scale during inference.

\begin{figure}[htbp]
\centerline{\includegraphics[scale=0.5]{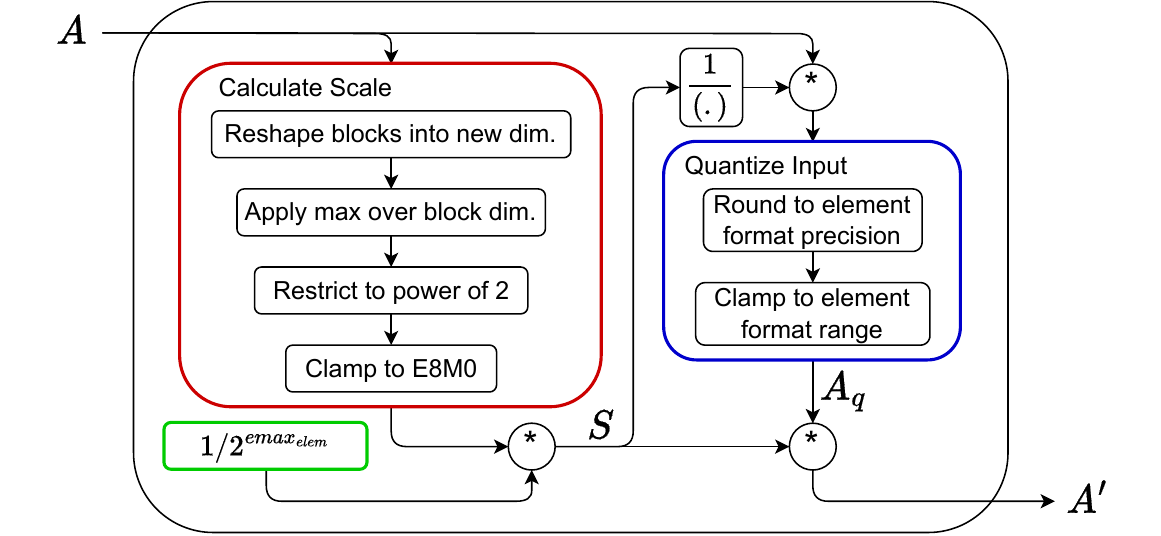}}
\caption{Features of our quantizer, each block here can be customised or replaced to implement other quantization schemes. The $2^{emax_{elem}}$ term refers to the largest exponent possible in the element format and $A'$ represents a real valued tensor formed by applying the scale on $A_q$.}
\label{fig:brevitas_quant_arith}
\end{figure}

Our implementation of MX scaling is another important addition to Brevitas. In the original Brevitas, scales are stored as tensors consisting of a single element (per-tensor) or $C$ elements (per-channel). This allows the scale to be applied using the element-wise multiplication operator in Pytorch where the scale is broadcast along the missing dimensions. In the case of MX scaling, the scale sharing granularity is finer than one scale per vector, which is incompatible with this element-wise multiplication and broadcasting. In the memory-efficient case, the length of the principal dimension of the scale tensor would be the number of shared scales per vector. However, this would be incompatible with the current element-wise multiplication with broadcasting. Our implementation preserves compatibility by making a scale tensor with the same dimensions as the input tensor, with shared scales repeated for each element that uses them. While this is inefficient for memory, the scale tensor can be compressed by removing repeated elements before deployment. Figure~\ref{fig:brevitas_quant_arith} shows the quantization process in our MX quantizer.

\subsection{Dot Products}

Equation~\ref{eqn:dp_conv} calculates a 2D convolution. As the boundaries between shared scales change, the sums over $c$, $h'$ and $w'$ are of interest because the boundaries determine where scales can be factored out of additions. If scales are factored out, error-free addition can be used (denoted by operator $\Sigma$), implemented efficiently using Kulisch accumulators~\cite{BMF}~\cite{kulisch2011} for all supported formats (Table~\ref{tab:mx_concrete}). If scales could not be factored out, normalisation is required before and after addition (denoted by operator $\Xi$). Full multipliers are denoted by $\times$ while multiplications that can be reduced to additions/shifts are denoted by $\otimes$. In the case of per-tensor scaling, scales can be completely factored out. With per-channel scaling, the sum over $c$ crosses block boundaries (Equation~\ref{eqn:dp_conv_pc}) while the scales can be factored out of sums over $h'$ and $w'$. Our implementation of MX scaling (Equation~\ref{eqn:dp_conv_mx}) sets the $C$ dimension as the {\em principal dimension}, because this is typically the longest dimension used for reduction during dot products in CNNs and gives more choices for $k$. This means that sums over $h'$ and $w'$ use normalisation while sums over $C$ use a mixture of both addition types. The ratio of normalisation to integer adders is controlled by the block size.

\begin{multline}
\label{eqn:dp_conv_pc}
\begin{aligned}
&A'_{n,h,w,l} = \sumi_{c=1}^{C} \ (s_{c} \otimes t_{c}) \\
&\otimes \sum_{h'=1}^{H'} \sum_{w'=1}^{W'} A_{q}(n, h+h', w+w', c) \times F_{q}(l, c, h', w')
\end{aligned}
\end{multline}

\section{Evaluation}

Our library of IP cores and our exploration infrastructure were evaluated on image classification on ImageNet using ResNet-18~\cite{He2015DeepRL}. A reference model was trained using FP32 and this pre-trained model was used as a starting point for all quantization schemes tested. MX formats were tested alongside per-tensor and per-channel schemes by applying quantization to all weight and activation tensors in the model. In all cases, the scale was restricted to E8M0 and computed at inference. The variables between schemes were the granularity of the shared scale and element format which was restricted to 8 bits or less per element as this is what the MX standard is aimed at. Synthesis was done using Vivado 2023.1, targetting a Xilinx Zynq UltraScale+ xczu7ev-ffvc1156-2-e device.

\begin{multline}
\label{eqn:dp_conv_mx}
\begin{aligned}
&A'_{n,,h,w,l} = \sumi_{c=1}^{\lceil \frac{C}{k} \rceil} \sumi_{h'=1}^{H'} \sumi_{w'=1}^{W'} \ (s_{l,c,h',w'} \otimes t_{l,c,h',w'}) \\
&\otimes \sum_{p=1}^{k} (A_{q}(n, h+h', w+w', c k+p) \\
&\quad \quad \quad \quad \times F_{q}(l, c k+p, h',w'))
\end{aligned}
\end{multline}

\subsection{Network-level Accuracy}

The accuracy of each quantization scheme was evaluated under both post-training quantization (PTQ) and quantization-aware training (QAT) using our exploration infrastructure introduced in Section~\ref{sec:explr_infra}. PTQ was performed by rounding the parameters of the FP32 model to the target scheme using "round to nearest even". Only linear and convolutional layers in the ResNet-18 model had quantization applied, all other operations such as batchnorm operations and ReLU activations continued to use single-precision~\cite{zhou2018dorefanet}, because convolutional layers and linear layers consist of more parameters and perform far more operations than the other layers. QAT was performed by rounding the reference model to a target scheme, then training further with a fine-tuning set.

A variety of MX formats have been evaluated, including all the concrete formats defined in the standard. The error on the test dataset after PTQ and QAT is plotted in Figure~\ref{fig:error_area}. For the MXINT family of formats, bit width has the largest impact on error, increasing bitwidth decreases error. Aside from that, block size also has an impact on error where decreasing block size decreases error. This is also reflected in the MXFP family, but block size has a smaller impact on accuracy.

\subsection{Core-level Hardware Results}

For rapid evaluation of MX formats, we create area models by profiling with out of context synthesis. The models are on a per IP block basis, for each of: multiplier arrays, adder trees and normalisation circuits. Multiplier arrays and FP adder trees have linearly increasing area with respect to the number of multipliers/adders. The coefficients in the linear model are found by least-squares fitting to a subset of synthesised designs. Normalisation circuits are similar, with a linear increase in area with the number of values to be normalised (number of output activations in a layer). For integer adder trees, our model calculates the sum of area of all adders in the tree. The logarithmic increase of adder size with tree depth is taken into account. The area for each individual adder and multiplier was found by synthesising while sweeping mantissa/exponent widths. This area model was used to estimate the area that would be required to unroll all of the linear and convolutional layers in the model, ignoring the cost of other operations as these will use a relatively small amount of area. Placing such an unrolled model on a single device is not feasible due to the large area required, however, this measure captures the effect of changing block size across a range of layers with varying $C$. The same model was also used to estimate the area of per-channel and per-tensor schemes.

As for latency and throughput, our individual cores are pipelined; there is no significant change in the clock frequency, and hence the core-level throughput achievable across the range of parameters we have explored. The same would be true for a fully-unrolled implementation. The number of pipeline stages does increase with $k$ in our implementation, due to the depth of the comparator tree as detailed in Section~\ref{sec:impl}. This effect was negligible in our experiments due to its logarithmic complexity and the small number of pipeline stages relative to other components. The number of pipeline stages in a FP32 implementation would be much larger as FP32 multipliers/adders (IP from Vivado) require more pipeline stages to match the frequency of our quantized implementation.

\begin{figure}[htbp]
\centerline{\includegraphics[scale=0.44]{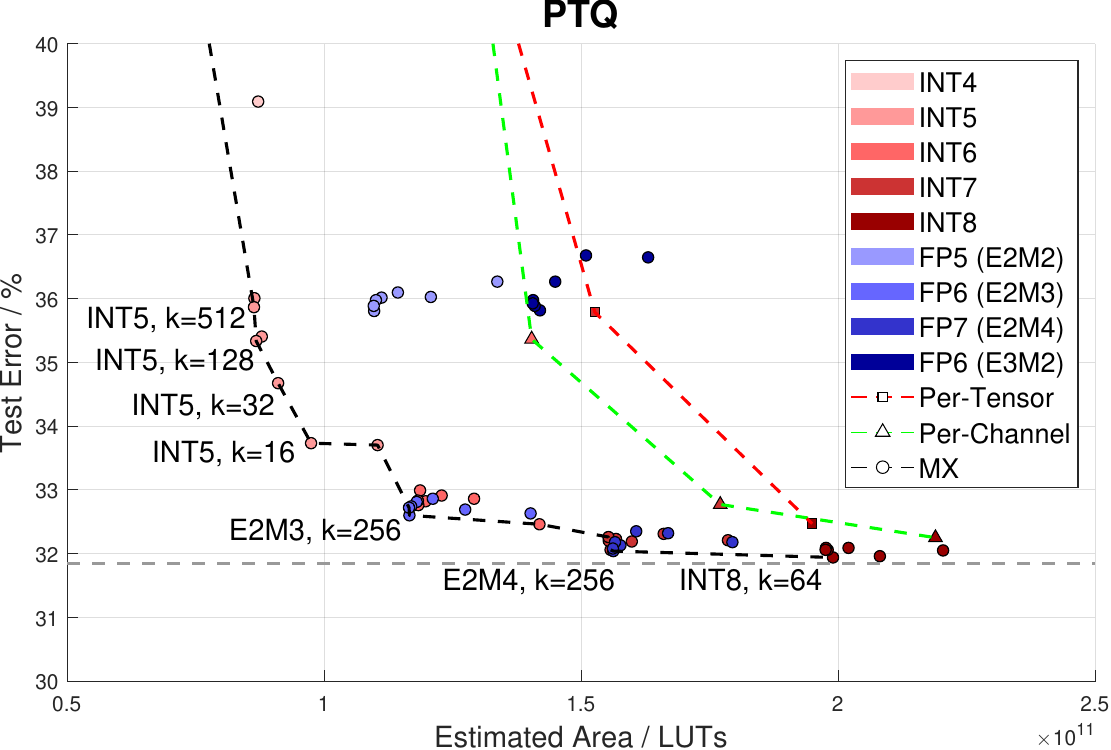}}
\vspace{0.3cm}
\centerline{\includegraphics[scale=0.44]{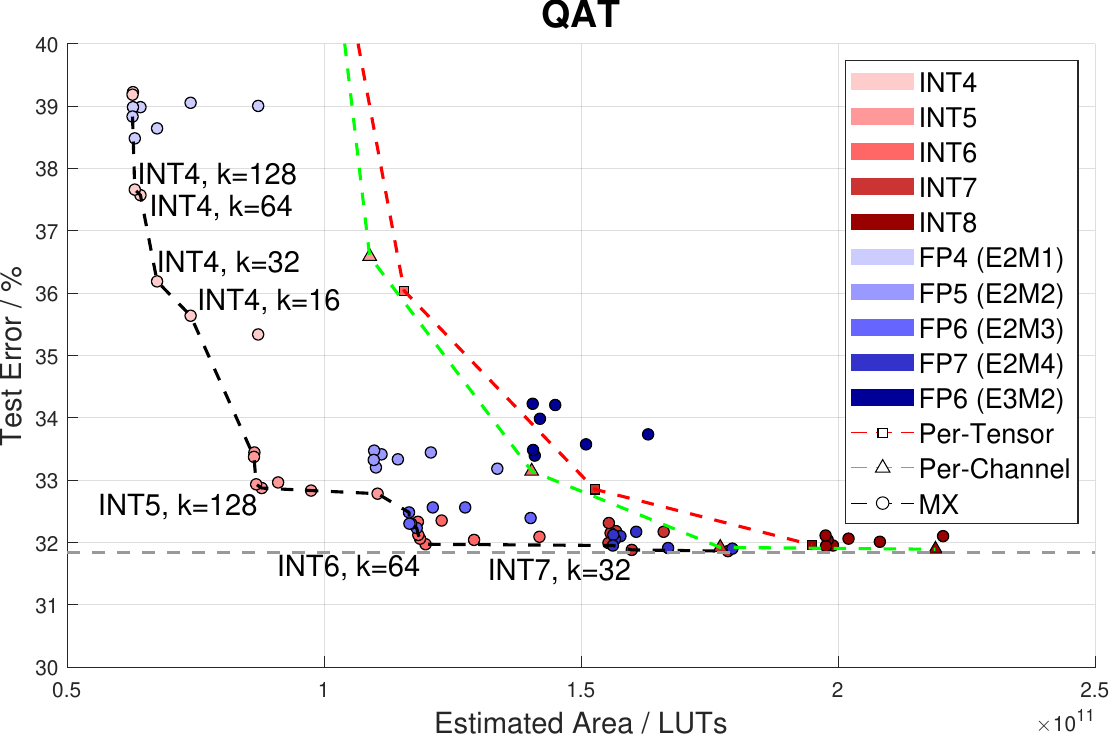}}
\caption{Error vs. estimated area of quantization schemes. Marker shape shows scale sharing regime. The grey dotted line is the FP32 baseline, other dotted lines show Pareto fronts. Pareto-optimal points are labelled with format and block size. Only schemes that offered more than 60\% accuracy are shown.}
\label{fig:error_area}
\end{figure}

Figure~\ref{fig:error_area} shows the test error and estimated area of quantization schemes before and after QAT. The plots omit schemes with more than 40\% error and schemes with high area utilization that provided little error improvement. The MXFP8 formats (E4M3/E5M2) are omitted because the error-free accumulation within our Dot implementation scales with $O(2^e)$ and uses large area. In our PTQ results, the Pareto optimal points which offer the lowest area cost are the MXINT5 formats. The MXINT6/7/8 formats with coarse grained scales provide accuracy close to the FP32 baseline. The MXFP6/7 formats E2M3 and E2M4 provide a marginal improvement over MXINT6/7. After QAT, the MXINT4 formats become the most desirable for low area, with MXINT5/6/7 formats providing near baseline accuracy. Generally across both PTQ and QAT results, if low area is desired (left side of plots), it is desirable to use a narrow MXINT format and the main design choice is the block size which can be used to trade area with accuracy. On the other hand if near baseline accuracy (grey dotted line) is desired, the wider MXINT formats are better suited, with bit width being the most impactful design choice.

The results also show the effect of our exploration infrastructure. Notably, QAT brings down the error cost of 4-bit formats such as MXFP4 (E2M1) and MXINT4 and makes them feasible for implementations with limited area. QAT also brings the MXINT5/6 formats to near baseline accuracy. However, the effect of QAT on MXFP formats was not as significant. MX formats considerably improve on the area/error tradeoff for FPGA implementation, compared to per-tensor and per-channel scaling, however in this application it is mainly the MXINT formats, i.e. block floating point, that provide the best results, rather than narrow-width MXFP.

\section{Conclusion}

In this paper, we have introduced an open-source MX compatible library of arithmetic components, which can be used to implement ML accelerator designs on FPGAs. Our library fully supports all concrete formats introduced by the MX standard, and is fully parameterised to support a wide range of element formats and block sizes beyond the concrete formats, too. Alongside this library, we have developed a software exploration infrastructure for MX which facilitates training and evaluating MX quantized models on GPUs. Our exploration infrastructure is fully integrated with Brevitas, allowing MX to be included in design space exploration alongside other traditional quantization schemes.

Finally, we explored the trade-off between inference accuracy and FPGA area for a range of formats introduced under the MX standard. Our findings show the benefit of using narrow formats such as MXINT4/5 and MXFP6/7 (E2M3, E2M4) over traditional per-tensor or per-channel quantization. Our experiments also show that MXINT6/7 are more desirable in this trade-off than the concrete MXINT8 format at times.

Our exploration infrastructure opens up a lot of interesting design choices. In the future, it could be used to explore mixed-precision models with different quantization schemes MX or non-MX. The scale also opens up some areas for exploration such as different scale formats (FP/INT) or other scale computation methods. Following this, IP Cores could be created for the best schemes from this exploration.

\bibliographystyle{IEEEtran}
\bibliography{ref}

\begin{thebibliography}{10}
\providecommand{\url}[1]{#1}
\csname url@samestyle\endcsname
\providecommand{\newblock}{\relax}
\providecommand{\bibinfo}[2]{#2}
\providecommand{\BIBentrySTDinterwordspacing}{\spaceskip=0pt\relax}
\providecommand{\BIBentryALTinterwordstretchfactor}{4}
\providecommand{\BIBentryALTinterwordspacing}{\spaceskip=\fontdimen2\font plus
\BIBentryALTinterwordstretchfactor\fontdimen3\font minus \fontdimen4\font\relax}
\providecommand{\BIBforeignlanguage}[2]{{%
\expandafter\ifx\csname l@#1\endcsname\relax
\typeout{** WARNING: IEEEtran.bst: No hyphenation pattern has been}%
\typeout{** loaded for the language `#1'. Using the pattern for}%
\typeout{** the default language instead.}%
\else
\language=\csname l@#1\endcsname
\fi
#2}}
\providecommand{\BIBdecl}{\relax}
\BIBdecl

\bibitem{Kaplan2020ScalingLF}
\BIBentryALTinterwordspacing
J.~Kaplan, S.~McCandlish, T.~J. Henighan, T.~B. Brown, B.~Chess, R.~Child, S.~Gray, A.~Radford, J.~Wu, and D.~Amodei, ``{Scaling Laws for Neural Language Models},'' \emph{ArXiv}, vol. abs/2001.08361, 2020. [Online]. Available: \url{https://api.semanticscholar.org/CorpusID:210861095}
\BIBentrySTDinterwordspacing

\bibitem{ocp_mx}
B.~D. Rouhani, N.~Garegrat, T.~Savell, R.~Zhao, A.~More, K.-N. Han, M.~Hall, J.~Klar, E.~Chung, Y.~Yu, M.~Schulte, R.~Wittig, I.~Bratt, N.~Stephens, J.~Milanovic, J.~Brothers, P.~Dubey, M.~Cornea, A.~Heinecke, M.~L. Andres~Rodriguez, S.~Deng, M.~Naumov, P.~Micikevicius, M.~Siu, and C.~Verrilli, ``{OCP Microscaling Formats (MX) Specification}.''

\bibitem{8766229}
IEEE, ``{IEEE Standard for Floating-Point Arithmetic},'' \emph{IEEE Std 754-2019 (Revision of IEEE 754-2008)}, pp. 1--84, 2019.

\bibitem{8280150}
R.~DiCecco, L.~Sun, and P.~Chow, ``{FPGA-based training of convolutional neural networks with a reduced precision floating-point library},'' in \emph{2017 International Conference on Field Programmable Technology (ICFPT)}, 2017, pp. 239--242.

\bibitem{Courbariaux2014TrainingDN}
\BIBentryALTinterwordspacing
M.~Courbariaux, Y.~Bengio, and J.-P. David, ``Training deep neural networks with low precision multiplications,'' \emph{arXiv: Learning}, 2014. [Online]. Available: \url{https://api.semanticscholar.org/CorpusID:16349374}
\BIBentrySTDinterwordspacing

\bibitem{Nagel2021AWP}
\BIBentryALTinterwordspacing
M.~Nagel, M.~Fournarakis, R.~A. Amjad, Y.~Bondarenko, M.~van Baalen, and T.~Blankevoort, ``{A White Paper on Neural Network Quantization},'' \emph{ArXiv}, vol. abs/2106.08295, 2021. [Online]. Available: \url{https://api.semanticscholar.org/CorpusID:235435934}
\BIBentrySTDinterwordspacing

\bibitem{Dai2021VSQuantPS}
\BIBentryALTinterwordspacing
S.~Dai, R.~Venkatesan, H.~Ren, B.~Zimmer, W.~J. Dally, and B.~Khailany, ``{VS-Quant: Per-vector Scaled Quantization for Accurate Low-Precision Neural Network Inference},'' \emph{MLSys 2021}, vol. abs/2102.04503, 2021. [Online]. Available: \url{https://api.semanticscholar.org/CorpusID:231855747}
\BIBentrySTDinterwordspacing

\bibitem{rouhani2023microscaling}
B.~D. Rouhani, R.~Zhao, A.~More, M.~Hall, A.~Khodamoradi, S.~Deng, D.~Choudhary, M.~Cornea, E.~Dellinger, K.~Denolf, S.~Dusan, V.~Elango, M.~Golub, A.~Heinecke, P.~James-Roxby, D.~Jani, G.~Kolhe, M.~Langhammer, A.~Li, L.~Melnick, M.~Mesmakhosroshahi, A.~Rodriguez, M.~Schulte, R.~Shafipour, L.~Shao, M.~Siu, P.~Dubey, P.~Micikevicius, M.~Naumov, C.~Verrilli, R.~Wittig, D.~Burger, and E.~Chung, ``{Microscaling Data Formats for Deep Learning},'' \emph{ArXiv}, 2023.

\bibitem{BMF}
S.~Fox, S.~Rasoulinezhad, J.~Faraone, D.~Boland, and P.~Leong, ``{A Block Minifloat Representation For Training Deep Neural Networks},'' in \emph{ICLR 2021}, 2021.

\bibitem{Zhang_2023}
\BIBentryALTinterwordspacing
C.~Zhang, J.~Cheng, I.~Shumailov, G.~Constantinides, and Y.~Zhao, ``{Revisiting Block-based Quantisation: What is Important for Sub-8-bit LLM Inference?}'' in \emph{Proceedings of the 2023 Conference on Empirical Methods in Natural Language Processing}.\hskip 1em plus 0.5em minus 0.4em\relax Association for Computational Linguistics, 2023. [Online]. Available: \url{http://dx.doi.org/10.18653/v1/2023.emnlp-main.617}
\BIBentrySTDinterwordspacing

\bibitem{ocp_fp8}
P.~Micikevicius, S.~Oberman, P.~Dubey, M.~Cornea, A.~Rodriguez, I.~Bratt, R.~Grisenthwaite, N.~Jouppi, C.~Chou, A.~Huffman, M.~Schulte, R.~Wittig, D.~Jani, and S.~Deng, ``{OCP 8-bit Floating Point Specification (OFP8)}.''

\bibitem{ip_cores}
\BIBentryALTinterwordspacing
E.~Samson, ``{MX-for-FPGA},'' 2023. [Online]. Available: \url{https://github.com/ebby-s/MX-for-FPGA}
\BIBentrySTDinterwordspacing

\bibitem{Kulisch+2012}
\BIBentryALTinterwordspacing
U.~Kulisch, \emph{{Computer Arithmetic and Validity}}.\hskip 1em plus 0.5em minus 0.4em\relax Berlin, Boston: De Gruyter, 2012. [Online]. Available: \url{https://doi.org/10.1515/9783110301793}
\BIBentrySTDinterwordspacing

\bibitem{doi:10.1137/1.9780898718027}
\BIBentryALTinterwordspacing
N.~J. Higham, \emph{{Accuracy and Stability of Numerical Algorithms}}, 2nd~ed.\hskip 1em plus 0.5em minus 0.4em\relax Society for Industrial and Applied Mathematics, 2002. [Online]. Available: \url{https://epubs.siam.org/doi/abs/10.1137/1.9780898718027}
\BIBentrySTDinterwordspacing

\bibitem{DBLP:journals/corr/abs-1903-08066}
\BIBentryALTinterwordspacing
S.~R. Jain, A.~Gural, M.~Wu, and C.~Dick, ``{Trained Uniform Quantization for Accurate and Efficient Neural Network Inference on Fixed-Point Hardware},'' \emph{CoRR}, vol. abs/1903.08066, 2019. [Online]. Available: \url{http://arxiv.org/abs/1903.08066}
\BIBentrySTDinterwordspacing

\bibitem{Choi2018PACTPC}
\BIBentryALTinterwordspacing
J.~Choi, Z.~Wang, S.~Venkataramani, P.~I.-J. Chuang, V.~Srinivasan, and K.~Gopalakrishnan, ``{PACT: Parameterized Clipping Activation for Quantized Neural Networks},'' \emph{ArXiv}, vol. abs/1805.06085, 2018. [Online]. Available: \url{https://api.semanticscholar.org/CorpusID:21721698}
\BIBentrySTDinterwordspacing

\bibitem{brevitas_mx}
\BIBentryALTinterwordspacing
E.~Samson, ``{Brevitas-MX},'' 2023. [Online]. Available: \url{https://github.com/ebby-s/brevitas}
\BIBentrySTDinterwordspacing

\bibitem{brevitas}
\BIBentryALTinterwordspacing
A.~Pappalardo, ``Xilinx/brevitas,'' 2023. [Online]. Available: \url{https://doi.org/10.5281/zenodo.3333552}
\BIBentrySTDinterwordspacing

\bibitem{mx_emu}
\BIBentryALTinterwordspacing
Microsoft, ``microsoft/microxcaling,'' 2023. [Online]. Available: \url{https://github.com/microsoft/microxcaling}
\BIBentrySTDinterwordspacing

\bibitem{kulisch2011}
\BIBentryALTinterwordspacing
U.~Kulisch, ``Very fast and exact accumulation of products,'' \emph{Computing}, vol.~91, no.~4, pp. 397--405, 2011. [Online]. Available: \url{https://doi.org/10.1007/s00607-010-0131-y}
\BIBentrySTDinterwordspacing

\bibitem{He2015DeepRL}
\BIBentryALTinterwordspacing
K.~He, X.~Zhang, S.~Ren, and J.~Sun, ``{Deep Residual Learning for Image Recognition},'' \emph{2016 IEEE Conference on Computer Vision and Pattern Recognition (CVPR)}, pp. 770--778, 2015. [Online]. Available: \url{https://api.semanticscholar.org/CorpusID:206594692}
\BIBentrySTDinterwordspacing

\bibitem{zhou2018dorefanet}
S.~Zhou, Y.~Wu, Z.~Ni, X.~Zhou, H.~Wen, and Y.~Zou, ``{DoReFa-Net: Training Low Bitwidth Convolutional Neural Networks with Low Bitwidth Gradients},'' 2018.

\end{thebibliography}

\end{document}